# بهینه‌سازی کدهای دودویی

پرویز قره‌باقری[1]، سیدحمید حاجی سیدجوادی[1]، پروانه اصغری[2*]، ناصر قره‌باقری[3]

*نویسنده مسئول

[1] دانشکده علوم، گروه ریاضی و علوم کامپیوتر، دانشگاه شاهد، تهران، ایران
[2] گروه مهندسی کامپیوتر، واحد تهران مرکزی، دانشگاه آزاد اسلامی، تهران، ایران
[3] دانشکده علوم، گروه ریاضی ، دانشگاه مراغه، آذربایجان شرقی، ایران

## چکیده

در این‌مقاله نشان داده شده است که می‌توان هر نوع داده دودویی را به‌صورت اجتماعی از کدکلمه‌هایی با طول متغیر تعریف کرد. این ویژگی به ما کمک می‌کند که بتوان نگاشتی یک‌به‌یک و پوشا از کدکلمه‌های پیشنهادی به کدکلمه‌های مورد نیاز تعریف کرد. از این‌رو با جایگزینی کدکلمه‌های جدید، داده‌های دودویی به داده‌های دودویی دیگری در راستای اهداف موردنظر تبدیل می‌گردد. یکی از این اهداف، کاستن حجم داده است. یعنی به‌جای کدکلمه‌های پیشنهادی هر داده‌ی دودویی، کدکلمه‌های هافمن را جایگزین نمود تا حجم داده کمتر گردد. یکی از ویژگی‌های این‌روش، نتیجه‌ی فشرده‌سازی مثبت برای هر نوع داده‌ی دودویی است، یعنی صرف‌نظر از حجم جدول کد، تفاضل حجم داده‌ی اصلی و حجم داده بعد از فشرده‌سازی، بزرگتر یا مساوی صفر خواهد شد. ویژگی مهم و کاربردی دیگر این‌روش، استفاده از کدکلمه‌های متقارن به‌جای کدکلمه‌های پیشنهادی به‌منظور ایجاد خواص تقارن، بازگشت پذیری و مقاومت در برابر خطا با قابلیت کدگشایی دوطرفه است.

## ۱- مقدمه

دنیای دیجیتال، دنیای صفر و یک‌هاست و مدیریت صفر و یک‌ها در فشرده‌سازی و انتقال داده بسیار مهم است. در این‌مقاله ادعا شده است که هر داده‌ی دودویی به‌صورت اجتماعی از کدکلمه‌ها[1]، قابل نمایش است. این‌مقاله سعی دارد این نگاه جدید را معرفی نموده و از آن بهره‌برداری مفید کند. یکی از کاربردهای این ایده‌ی پیشنهادی، کم کردن حجم داده و ارسال سریع‌تر آن‌هاست. بر این اساس سعی شده است داده‌های دودویی یک فایل به اجتماعی از کدکلمه‌ها دسته‌بندی شده و با کدکلمه‌های هافمن جایگزین شوند. این ایده علاوه بر کاستن حجم داده می‌تواند در متقارن‌سازی بیت‌ها به منظور مقاومت در برابر خطا مفید واقع شود. در ادامه جهت آشنایی بیشتر با الگوریتم‌های کدگذاری، مروری بر تلاش‌هایی که تا کنون در راستای کدگذاری و فشرده‌سازی داده انجام گرفته شده است خواهد شد.

ریچارد همینگ برای اولین‌بار در سال ۱۹۴۸ نظریه‌ی کدگذاری خود را پایه‌ریزی کرد. در همین حین، کلود شانون نیز نظریه‌ی کدگذاری بدون نویز و پس از آن نظریه آنتروپی خود را ارائه نمود که باعث رشد نظریه اطلاعات گردید. امروزه روش کدگذاری شانون یک روش پایه محسوب می‌گردد[۱, ۴, ۶].

در سال ۱۹۵۱ میلادی هافمن، به کمک ایده‌ی استفاده از درخت دودیی مرتب شده بر حسب تکرار و بر اساس اصل مفهوم آنتروپی تعریف شده توسط شانون، توانست کدگذاری هافمن را ابداع و اثبات کند[۱۱، ۱۳]. کدهای هافمن تنها کدهای بهینه‌ی مبتنی بر آنتروپی می‌باشد که توسط وی در سال ۱۹۵۲ منتشر شد[۵]. این‌روش با فرض دریافت آرایه‌ای مرتب شده از سمبل‌های یک منبع ناصفر، می‌تواند در زمان $O(nlogn)$ ، سمبل‌های دریافتی را در حالت بهینه کدگذاری کند. اگرچه شانون پیش از هافمن کدهای خود را معرفی نمود ولی کدهایش به دلیل عدم بهینگی نتوانست مانند کدهای هافمن فراگیر شود ولی بعد از وی تلاش‌هایی جهت نزدیک کردن آن به بهینگی گرفت[۲, ۳]. کدهای گولومب[2] که بعد از هافمن معرفی گردید اگرچه دارای مرتبه زمانی خطی بود ولی به دلیل عدم بهینگی، بیشتر برای کدگذاری حالاتی خاص از فرکانس‌های ورودی

کاربرد پیدا کرد که از جمله‌ی آن می‌توان، کدگذاری فرمت ویدئویی H.264/AVC را نام برد[۷]. الگوریتم‌های ام-تی-اف و بی-دبلیو-تی نیز، هزینه‌ی بهتری نسبت به روش‌های پیشین در کدگذاری و کدگشایی خود ارائه ندادند. از کدگذاری‌های دیگر، روش کدگذاری شمارشی منبع[۸] می‌باشد که در سال ۱۹۷۳ ابداع و در سال ۱۹۸۴ در یکی مقالات منتشر شده از این‌روش به منظور فشرده‌سازی استفاده شد. همچنین می‌توان به کدگذاری گرامری که بر مبنای نظریه‌ی زبان‌ها و ماشین‌ها پایه‌ریزی و پیاده‌سازی شده است اشاره کرد. در سال ۲۰۱۰ نیز روش متفاوت کدگذاری منبع بدون اتلاف قطبی[۹] با زمان کدگذاری و کدگشایی غیرخطی منتشر گردید. همچنین مقالاتی نیز به کاهش داده های باینری با توجه به کدهای شانون پرداخته اند[۱۲، ۱۴، ۱۵، ۱۶] مقالاتی نیز توسط تئوری های ارائه شده به کم کردن طول بیتی داده ها متمرکز شده اند[۴، ۱۴]. همچنین می‌توان فشرده‌سازی به روش ال‌زد۷۷[۳][۱] را نیز روشی موفق در عرصه‌ی فشرده‌سازی بدون اتلاف داده مخصوصا در تصاویر است را نام برد که در مواردی بهتر از الگوریتم‌های هافمن و شانون عمل کرده است. از الگوریتم‌های مبتنی بر دیکشنری مانند الگوریتم‌های خانواده ال‌زدبلیو در فشرده‌سازی با اتلاف تصاویر نیز استفاده شده است[۱۰].

آنچه مهم است ابداع الگوریتمی جدید برای بهره‌گیری بهتر و بیشتر از الگوریتم هافمن می‌باشد. همانطور که گفته شد الگوریتم پیشنهادی برای بهینه‌سازی هر نوع داده‌ی بیتی که توسط هر نوع الگوریتمی کدگذاری شده باشد طراحی و پیاده‌سازی شده است و نتایج آن بر روی داده‌های دودویی تصادفی نیز درست است. از نتایج این‌روش در این‌مقاله فشرده‌سازی ۵ الی ۱۰ درصدی داده‌های دودویی است.

آنچه در این‌مقاله به‌عنوان نوآوری اصلی بیان شده است طرح این‌ادعاست که هر داده‌ی دودویی به‌صورت اجتماعی از کدکلمه‌هایی با طول متغیر قابل نمایش است. این‌ادعا بر طرح یک مساله‌ی کاربردی در کدگذاری اشاره دارد که در ظاهر ساده به نظر می‌رسد ولی می‌تواند کاربردهای فراوانی از جمله فشرده‌سازی و متقارن‌سازی در کدگذاری داشته باشد. همچنین ادعای نتیجه‌ی مثبت فشرده‌سازی برای هر نوع داده‌ی دودویی ورودی ادعایی است که به‌صورت خاص در این‌مقاله به عنوان نوآوری به آن توجه شده است. همچنین همانطور که نشان داده خواهد شد برای اثبات این‌ادعاها از یک لم و دو قضیه‌ی پیشنهادی استفاده شده است.

در ادامه و در فصل دوم، الگوریتم‌های پایه، معرفی و در فصل سوم الگوریتم پیشنهادی تشریح می‌گردد. در فصل چهارم ادعای فشرده‌سازی مثبت مطرح شده و فصل پنجم به نتایج حاصل از پیاده‌سازی نرم افزاری الگوریتم می‌پردازد. در فصل ششم نیز کارهای آتی و نتیجه‌گیری گنجانده شده است.

## ۲- مروری اجمالی بر الگوریتم‌های پایه کدگذاری

### ۲-۱- الگوریتم هافمن

الگوریتم هافمن[۱، ۱۱، ۱۳] یک روش کدگذاری بهینه‌ی منبع می‌باشد که دارای ۳ شرط اساسی در کدگذاری خود می‌باشد:

۱-احتمال حروف با طول کد منتسب به آن رابطه‌ی عکس دارد.

۲-دو حرف با پایین ترین احتمال، کدی با طول یکسان خواهند داشت.

۳-دو تا از کم احتمال ترین حروف، جز در بیت آخر دارای کدهای یکسان می‌باشند .

الگوریتم هافمن به‌صورت الگوریتم زیر بیان می‌شود:

۱-متن مورد نظر را دریافت کن.

۲-احتمال تکرار k امین حرف در متن را طبق رابطه‌ی $P_k = \frac{n_k}{n}$ بدست آور.

۳-دو حرف با کمترین احتمال را دو گره در نظر گرفته و با هم ترکیب کن و حاصل این ترکیب را گره جدید در نظر بگیر و این مرحله را تا آنجا که تنها یک گره بدون یال باقی بماند تکرار کن.

۴-از گره آخر تا رسیدن به گره‌های اولیه، شروع کرده و در جهت عکس حرکت کن. برای هر گره، به یکی از یال‌هایی که به سمت آن آمده عدد "۰" و به دیگری عدد "۱" را نسبت بده.

۵-اکنون جهت بدست آمدن کد هافمن حرف مورد نظر، از گره انتخابی تا گره اولیه در مسیر معکوس حرکت کرده و مقادیر یال‌های خوانده شده را کنار هم بگذار.

جهت روشن شدن مبحث فرض کنید در یک متن که از ۵ حرف(سمبل) تشکیل شده است احتمال حروف به‌صورت p(a)=0.4, p(b)=0.2, p(c)=0.2, p(d)=0.1, p(e)=0.1 باشد طبق الگوریتم داریم:

0.1 + 0.1=0.2
0.2 +0.2=0.4
0.2 +0.4=0.6
0.4 + 0.6=1

از پایین به احتمال 0.6 کد "۱" و به احتمال 0.4 کد "۰" را اختصاص می‌دهیم این روند را برای بقیه احتمالات تکرار می‌کنیم با این تفاوت که کدهای قبلی در پشت کد احتمالات بعدی به ارث برده می‌شود یعنی به احتمال 0.4 بعدی، کد "۱۱" و برای احتمال 0.2 کد "۱۰" تعلق می‌گیرد. می‌توان این ساختار را به کمک درخت بهتر نمایش داد. برای بدست آوردن کد هافمن هر حرف، درخت احتمال بر اساس الگوریتم گفته شده، رسم شده و کدهای ۰ و ۱ از بالا به پایین تا رسیدن به برگی که احتمال حرف مورد نظر را دارد خوانده می‌شود.

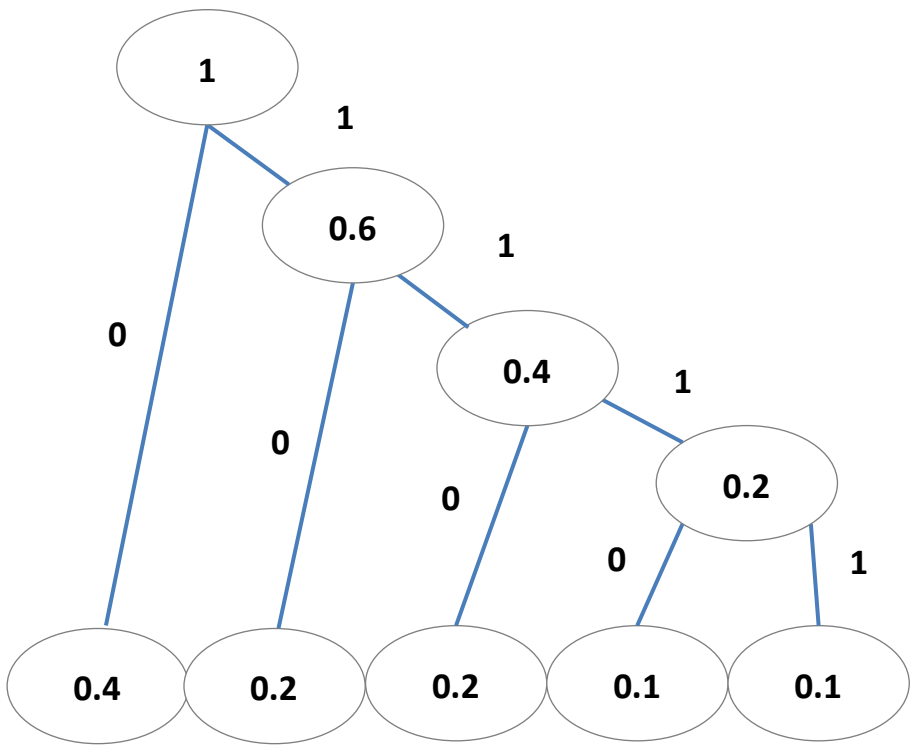

شکل ۱- درخت هافمن

بنابراین برای بدست آوردن کدهای هافمن بر اساس شکل۱ خواهیم داشت: a="0", b="10", c="110", d="1110", d="1111" . همانطور که نتایج نشان می‌دهد کدکلمه‌های هافمن برای کاراکترهای پرتکرار طول کوتاه‌تری خواهد داشت.

### ۲-۲- الگوریتم شانون

این‌روش نیز مشابه روش هافمن می‌باشد که نوع کدگذاری درختی آن بر خلاف روش هافمن که از پایین به بالا در نظر گرفته شده است، از بالا به پایین طراحی گردیده و نتایج کدگذاری آن در بهترین حالت برابر روش هافمن است [۶، ۱۳].

در کدگذاری شانون، سمبل‌ها پس از مرتب‌سازی بر اساس احتمالاتشان، با نمایش دودویی عدد حاصل از مجموع احتمالات سمبل‌های قبل از خودشان(احتمال تجمعی $(\sum_{k=1}^{i-1} p_k)$)، به‌طول $l_i = \lceil -\log p_i \rceil$ بیت، کد می‌شوند. در اینجا $\lceil x \rceil$ بیانگر تابعی است که x را رو به بالا گرد می‌کند.

فرض کنید در یک‌متن که از ۵ حرف(سمبل) تشکیل شده است احتمال حروف به‌صورت p(a)=0.5, p(b)=0.25, p(c)=0.125, p(d)=0.063, p(e)=0.062 باشد بنابراین خواهیم داشت:

| **S** | **p(S)** | $l_i$ | $\sum_{k=1}^{i-1} p_k$ | **binary** | **Codeword** |
|---|---|---|---|---|---|
| a | 0.5 | 1 | 0.0 | 0.00000 | 0 |
| b | 0.25 | 2 | 0.5 | 0.10000 | 10 |
| c | 0.125 | 3 | 0.75 | 0.11000 | 110 |
| d | 0.063 | 4 | 0.875 | 0.11100 | 1110 |
| e | 0.062 | 5 | 0.938 | 0.11110 | 11110 |

با توجه به کدکلمه‌های ساخته شده نتیجه می‌گیریم که کدهای شانون نوعی کدگذاری با طول متغیر می‌باشد و در حالتی از الگوریتم، کدهای استخراجی به‌صورت (حالت اول) ...۱،۰۱،۰۰۱،۰۰۰۱ یا مانند مثال، به‌صورت (حالت دوم) ....۰،۱۰،۱۱۰،۱۱۱۰ می‌باشند که تقریبا کدهای قابل پیش‌بینی نیز هستند. بنابراین به دلیل تشابه کدکلمه‌های با فرم $1^*0$ و فرم $0^*1$ با کدکلمه‌های شانون، در این‌مقاله از کدکلمه‌های پیشنهادی، باعنوان کدکلمه‌های شانونی نیز استفاده خواهد شد.

## ۳- الگوریتم پیشنهادی

با توجه به کدهای بدست‌آمده در فصل دوم پیداست که هر داده‌ی کدشده با کدکلمه‌های شانون در حالت اول را می‌توان به‌صورت $\{1^*0\}^+$ و حالت دوم را به‌صورت $\{0^*1\}^+$ نمایش داد. در ادامه از تعاریف، لم و قضایای پیشنهادی جهت اثبات ادعا استفاده خواهد شد.

تعریف۱: به هر داده‌ی دودویی که یک بیت"۱" در انتهای آن الحاق گردد داده دودویی الحاقی۱ گفته می‌شود.

تعریف۲: به هر داده‌ی دودویی که یک بیت"۰" در انتهای آن الحاق گردد داده دودویی الحاقی۰ گفته می‌شود.

لم۱: هر داده‌ی دودویی الحاقی۱ به‌صورت$\{1^*0\}^+$ و هر داده‌ی دودویی الحاقی۰ به‌صورت$\{0^*1\}^+$ که مجموعه‌ای از کدکلمه‌های مستقل است قابل نمایش است.

اثبات: برای اثبات این‌ادعا، داده‌ی دودویی الحاقی۱ را از چپ به راست می‌خوانیم، اگر با بیت"۱" شروع شود آن‌را به عنوان یک کدکلمه ثبت می‌کنیم و اگر با بیت"۰" شروع شود آن بیت را خوانده و از آن عبور می‌کنیم تا به بیت"۱" برسیم، مجموعه بیت‌های خوانده‌شده که به‌صورت $1^*0$ می‌باشد را به عنوان کلمه‌ی جدید می‌شناسیم. اگر این روند را تا انتها پیش ببریم متوجه خواهیم شد که هر داده‌ی دودویی به‌صورت $\{1^*0\}^+$ قابل نمایش است. همین اثبات برای هر داده‌ی دودویی الحاقی۰ نیز برقرار است با این تفاوت که اگر با بیت"۰" شروع شود آن‌را به عنوان یک کدکلمه ثبت می‌کنیم و اگر با بیت"۱" شروع شود آن بیت را خوانده و از آن عبور می‌کنیم تا به بیت"۰" برسیم . ■

با توجه به این ویژگی، می‌توان داده‌های دودویی را به‌صورت اجتماع کدکلمه‌ها معرفی کرد و فراوانی کدکلمه‌های موجود در آن‌را محاسبه نمود، سپس توسط الگوریتم هافمن کدگذاری کرده و جایگزین نمود. این الگوریتم به‌صورت زیر نشان داده شده است:

۱-داده را بگیر.

۲-بیت"۱" را به انتهای داده الحاق کن.

۳-از داده‌ی الحاقی۱ کدکلمه‌های به فرم $1^*0$ و از داده‌ی الحاقی۰ کدکلمه‌های به فرم $0^*1$است را جدا کرده و فراوانی هر کدکلمه را محاسبه کن.

۴-بر روی کدکلمه‌ها، الگوریتم هافمن را اجرا کرده و به‌جای آن‌ها، کدکلمه‌های هافمن را جایگزین کن.

۵-پایان

مرتبه‌ی زمانی الگوریتم پیشنهادی به‌صورت زیر قابل محاسبه است:

با توجه به اینکه مرتبه‌ی زمانی کدگذاری الگوریتم هافمن $O(nlogn)$ می‌باشد بنابراین خواهیم داشت:

$$O(Proposed\ ) = O(nh) + O(nlogn) = O(nlogn) \qquad (۱)$$

که در آن h، طول بزرگترین‌کدکلمه و n، تعداد کدکلمه‌هایی است که از اجتماع آن‌ها، داده‌ی دودویی شکل می‌گیرد. در ادامه با ذکر مثال۱ به پیاده‌سازی الگوریتم پیشنهادی پرداخته می‌شود. کدهای دودویی زیر را در نظر بگیرید:

۰۰۰۱۱۰۱۰۱۰۱۱۰۱۰۱۰۱۰۰۰۰۰۱۱۰۱۰۱۰۱۰۰۰۰۰۱۰۱۰۱۰۱۰۰۰۱۰۰۱۱۱
۰۰۱۰۱۰۰۰۰۰۰۱۰۱۰۱۰۱۰۰۰۰۱۰۱۰۱۰۰۱۰۱۰۱۰

پس از الحاق بیت"۱" به انتهای آن و تبدیل آن به کدهای الحاقی۱ ترتیب فراوانی کدکلمه‌های شانون موجود در آن به‌صورت زیر خواهد بود:

۰۰۰۱۱۰۱۰۱۰۱۱۰۱۰۱۰۱۰۰۰۰۰۱۱۰۱۰۱۰۱۰۰۰۰۰۱۰۱۰۱۰۱۰۰۰۱۰۰۱۱۱
۰۰۱۰۱۰۰۰۰۰۰۱۰۱۰۱۰۱۰۰۰۰۱۰۱۰۱۰۰۱۰۱۰۱۰۱

جدول ۱- جدول فراوانی کد شانون حالت اول

| کدکلمه‌های شانون | فراوانی |
|---|---|
| ۰۱ | ۲۱ |
| ۱ | ۵ |
| ۰۰۱ | ۳ |
| ۰۰۰۰۰۱ | ۳ |
| ۰۰۰۱ | ۲ |
| ۰۰۰۰۱ | ۱ |

حال با توجه به فراوانی کدها، آنها را توسط الگوریتم هافمن کدگذاری می‌کنیم.

جدول۲- کد هافمن به جای شانون حالت اول

| کدکلمه‌های شانون | فراوانی | کدکلمه‌های هافمن |
|---|---|---|
| ۰۱ | ۲۱ | ۰ |
| ۱ | ۵ | ۱۰۰ |
| ۰۰۱ | ۳ | ۱۰۱ |
| ۰۰۰۰۰۱ | ۳ | ۱۱۰ |
| ۰۰۰۱ | ۲ | ۱۱۱۰ |
| ۰۰۰۰۱ | ۱ | ۱۱۱۱ |

طول کد اولیه برابر ۸۶ بیت است. بعد از جایگذاری خواهیم داشت:

۱۱۱۰۱۰۰۰۰۰۰۱۰۰۰۰۰۱۱۰۱۰۰۰۰۰۰۱۱۰۰۰۰۱۱۱۰۱۰۱۱۰۰۱۰۰۱۰۱۰۱۱۰۰
۰۰۱۱۱۱۰۰۱۰۱۰۰۰

بنابراین طول کد خروجی برابر با ۶۶ بیت، و طول کد اصلی ۸۶ بیت می‌باشد که توسط الگوریتم پیشنهادی ۲۴ درصد از طول کد، کاهش یافته است و این یعنی فشرده‌سازی ۲۴ درصدی.

در اینجا کدهای خروجی را با الحاق یک بیت"۱" به الحاقی۱ تبدیل کرده و مجددا با الگوریتم پیشنهادی کدگذاری می‌گردد تا به طرح ادعایی جدید در فصل چهارم پرداخته شود.

۱۱۱۰۱۰۰۰۰۰۰۱۰۰۰۰۰۱۱۰۱۰۰۰۰۰۰۱۱۰۰۰۰۱۱۱۰۱۰۱۱۰۰۱۰۰۱۰۱۰۱۱۰۰
۰۰۱۱۱۱۰۰۱۰۱۰۰۰۱

جدول۲- کدگذاری در حالت اول

| کدکلمه‌های شانون | فراوانی | کدکلمه‌های هافمن |
|---|---|---|
| ۱ | ۱۲ | ۱ |
| ۰۱ | ۷ | ۰۰ |
| ۰۰۱ | ۳ | ۰۱۱۱ |
| ۰۰۰۱ | ۳ | ۰۱۱۰ |
| ۰۰۰۰۰۱ | ۳ | ۰۱۰۱ |
| ۰۰۰۰۱ | ۲ | ۰۱۰۰ |

با توجه به جدول کدگذاری مشخص است که بعد از کدگذاری مجدد طول کد به ۶۱ بیت کاهش می‌یابد و این نشان‌دهنده‌ی این است که این الگوریتم پیشنهادی قابلیت فشرده‌سازی مجدد داده‌های فشرده را دارد که در فصل چهارم اثبات خواهد شد.

در صورتی‌که کدکلمه‌های شانونی به‌صورت حالت دوم در نظر گرفته شود تقسیم بندی تغییر خواهد کرد.

داده اصلی را مجدد در نظر بگیرید که با الحاق بیت"۰" به داده‌ی دودویی الحاقی ۰ تبدیل می‌شود.

۰۰۰۱۱۰۱۰۱۰۱۱۰۱۰۱۰۱۰۰۰۰۰۱۱۰۱۰۱۰۱۰۰۰۰۰۱۰۱۰۱۰۱۰۰۱۰۰۱۱۱
۰۰۱۰۱۰۰۰۰۰۰۱۰۱۰۱۰۱۰۰۰۰۱۰۱۰۱۰۰۱۰۱۰۱۰

بنابر این کد اصلی به‌صورت زیر خواهد شد:

۰۰۰۱۱۰۱۰۱۰۱۱۰۱۰۱۰۱۰۰۰۰۰۱۱۰۱۰۱۰۱۰۰۰۰۰۱۰۱۰۱۰۱۰۰۱۰۰۱۱۱
۰۰۱۰۱۰۰۰۰۰۰۱۰۱۰۱۰۱۰۰۰۰۱۰۱۰۱۰۰۱۰۱۰۱۰۰

جدول۳- جدول فراوانی کد شانون حالت دوم

| کدکلمه‌های شانون | فراوانی |
|---|---|
| ۱۰ | ۲۵ |
| ۰ | ۲۴ |
| ۱۱۰ | ۳ |
| ۱۱۱۰ | ۱ |

حال با توجه به فراوانی کدها، آنها را توسط الگوریتم هافمن کدگذاری می‌کنیم.

جدول۴- کد هافمن به جای شانون حالت دوم

| کدکلمه‌های شانون | فراوانی | کدکلمه‌های هافمن |
|---|---|---|
| ۱۰ | ۲۵ | ۰ |
| ۰ | ۲۴ | ۱۰ |
| ۱۱۰ | ۳ | ۱۱۰ |
| ۱۱۱۰ | ۱ | ۱۱۱ |

در صورتی‌که جایگذاری انجام شود طول کد برابر خواهد شد با:

$$Len = 26 \times 1 + 23 \times 2 + 3 \times 30 + 1 \times 3 = 84$$

بنابراین میزان فشرده‌سازی ۳ درصدی خواهیم داشت که با توجه به فشرده‌سازی ۲۴ درصدی می توان دو نکته‌ی زیر را بیان کرد:

انتخاب حالت بهتر، منجر به فشرده‌سازی بیشتر خواهد شد.

نتیجه‌ی فشرده‌سازی در هر حالتی بزرگتر و مساوی صفر خواهد شد و این نتیجه‌ی مثبت، به نوع داده‌ی دودویی ارتباطی ندارد که در بخش(۴) به اثبات آن خواهیم پرداخت.

## ۴- اثبات فشرده‌سازی مثبت الگوریتم

در این بخش، با فرض تعاریف بخش(۳)، دو قضیه درباره‌ی فشرده‌سازی مثبت، پیشنهاد و اثبات می‌گردد بدین منظور و جهت بیان بهتر قضیه، داده‌ی دودویی اولیه با d، تابع کدگذار پیشنهادی با f و تعداد بیت‌های داده اولیه با $n(d)$ و تعداد بیت‌های داده بعد از الگوریتم کدگذاری پیشنهادی با$n(f(d))$ نمایش داده می‌شود.

قضیه۱. به ازای هر داده‌ی دودویی d که به عنوان ورودی تابع کدگذار f در نظر گرفته شود همواره تفاضل تعداد بیت‌های داده‌ی خروجی از تعداد بیت‌های داده‌ی ورودی بزرگتر یا مساوی صفر است یعنی:

$$\forall d \in \{0.1\}^* \qquad n(d) - n\big(f(d)\big) \geq 0. \qquad (۲)$$

اثبات:

(۱) می‌دانیم میانگین طول کدهای هافمن در بدترین حالت برابر میانگین طول کدهای شانون است[۱].

(۲) همچنین طبق لم۱ نتیجه می‌گیریم هر داده‌ی دودویی مجموعه‌ای از کدکلمه‌های شانون است. همانطور که تعریف شد طول بیت‌های داده‌ی دودویی را $n(d)$ در نظر می‌گیریم. همچنین بعد از اجرای الگوریتم پیشنهادی، کدکلمه‌های هافمن جایگزین کدکلمه‌های شانون می‌شود که تعداد بیت‌های داده باینری را بعد از جایگذاری، $n(f(d))$ نشان می‌دهیم. بنابراین بر اساس (۱) و (۲) خواهیم داشت $n(f(d)) \leq n(d)$ و این یعنی $n(d) - n\big(f(d)\big) \geq 0$ ■

این بدان معنی است که با صرف‌نظر کردن از حجم جدول کد، میزان فشرده‌سازی، همواره بزرگتر یا مساوی صفر خواهد بود.

قضیه۲. اگر داده‌ی دودویی d توسط تابع کدگذار f، به تعداد متناهی n مرتبه کدگذاری شود آنگاه : $n(f^n(d)) \leq n(f^{n-1}(d)) \qquad \forall d \in \{0.1\}^*$.

اثبات: با توجه به اینکه این قضیه برای هر نوع داده‌ی ورودی درست فرض شده است بنابراین اگر داده‌ی ورودی$f(d)$ فرض شود بنا بر قضیه۱ می‌توان ادعا کرد: $n(f(f(d))) \leq n(f(d))$. از این‌رو اگر این روال n بار تکرار شود خواهیم داشت:

$$n(f^n(d)) \leq n(f^{n-1}(d)) \qquad (۳)$$

بنابراین حکم ثابت است ■

## ۵- پیاده‌سازی الگوریتم پیشنهادی

الگوریتم پیشنهادی به کمک زبان بیسیک پیاده‌سازی گردید. در این پیاده‌سازی ابتدا در فایل‌های مورد آزمایش، کدکلمه‌های پیشنهادی شناسایی و فراوانی‌شان محاسبه شد. در مرحله‌ی بعد، این کدکلمه‌ها متناسب با فراوانی‌شان، با الگوریتم هافمن کدگذاری گردیده و کدکلمه‌های هافمن، جایگزین کدکلمه‌های پیشنهادی شد.

پس از پیاده‌سازی الگوریتم و اجرای یک مرحله از آن بر روی داده‌هایی از تصاویر رنگی پیچیده از جمله تصویر ‘‘Lena’’، داده‌های صوتی، متنی و داده‌های زیپ‌شده با نوع انتخاب تصادفی از بین فایل‌های شخصی، به‌طور میانگین نتیجه‌ی فشرده‌سازی ۵ درصدی بدست آمد.

## ۶- نتیجه‌گیری و کارهای آینده

در این‌مقاله نشان داده شد که هر داده‌ی دودویی را می‌توان به‌صورت اجتماعی از کدکلمه‌ها نوشت و این دیدگاه در جهت تبدیل یک داده به داده‌ی دیگر با قابلیت معکوس‌پذیری به‌منظور کاهش حجم مورد استفاده قرار گرفت. همچنین نشان داده شد که به‌کمک الگوریتم پیشنهادی، امکان فشرده‌سازی برای هر نوع داده‌ی دودویی، همچون داده‌های فشرده‌شده و حتی داده‌های تصادفی وجود دارد ولی هیچگاه ممکن نیست عکس فشرده‌سازی در آنها اتفاق بیفتد. زیرا ثابت شد میزان فشرده‌سازی همواره مثبت است و بر روی هر نوع داده‌ی دودویی قابل پیاده‌سازی و اجرا می‌باشد.

همچنین از ایده‌ی مطرح شده در این‌مقاله می‌توان پژوهش‌هایی را برای کارهای آتی پیشنهاد داد:

۱-مقاوم‌سازی داده‌ها در برابر خطا

در این‌روش به‌جای کدکلمه‌های پیشنهادی، کدکلمه‌های مقاوم در برابر خطا استفاده می‌شود.

۲-بازگشت‌پذیری و افزایش سرعت کدگشایی

در این‌روش نیز به‌جای کدکلمه‌های پیشنهادی، کدکلمه‌های متقارن استفاده می‌شود. از آنجاکه کدکلمه‌های متقارن، قابلیت مقاومت در برابر خطا را نیز دارد بنابراین برای این‌روش توضیحات بیشتری ارائه می‌گردد.

کدهای دودویی در اکثر روش‌های کدگذاری از جمله کدگذاری هافمن با کدکلمه‌هایی نامتقارن کد می‌شوند این کدها در برابر خطا مقاوم نبوده و فقط از یک‌طرف کدگشایی می‌شوند. در چنین کدهایی می‌توان از الگوریتم پیشنهادی جهت متقارن کردن کدها استفاده کرد. کدهای متقارن از کدکلمه‌هایی تشکیل شده‌اند که دارای تقارن بوده و دارای ویژگی‌هایی چون بازگشت پذیری، کدگشایی دوطرفه و مقاومت در برابر خطا می‌باشند. این مبحث مقاله‌ای مجزا را می‌طلبد ولی توضیحاتی مقدماتی از آن ارائه می‌گردد.

کدکلمه‌های متقارن زیر را در نظر بگیرید:

{0, 11, 101, 1001, 10001, …}

با توجه به جدول۵ می‌توان جدول فراوانی۶ را برای کدکلمه‌های متقارن نیز تعریف کرد.

جدول۶- کدکلمه‌های متقارن به جای شانون حالت دوم

| کدکلمه‌های شانون | فراوانی | کدکلمه‌های متقارن |
|---|---|---|
| ۱۰ | ۲۵ | ۰ |
| ۰ | ۲۰ | ۱۱ |
| ۱۱۰ | ۳ | ۱۰۱ |
| ۱۱۱۰ | ۱ | ۱۰۰۱ |

با جایگذاری خواهیم داشت:

۱۰۱۱۱۱۰۱۰۰۰۱۱۱۱۱۱۱۱۱۰۱۰۰۰۱۱۱۱۱۱۱۱۰۰۰۰۱۱۱۱۰۱۱۱۰۰۱۱۱۰۰
۱۱۱۱۱۱۱۱۰۰۰۰۱۱۱۱۱۱۱۰۰۰۱۱۰۰۰

طول‌کد برابر ۷۸ بیت می‌باشد. اگرچه جایگذاری کدهای متقارن نمی‌تواند طول کوتاه‌تری نسبت به کدهای هافمن ایجاد کند ولی همانطور که گفته شد این کدها متقارن بوده و دارای ویژگی‌هایی چون بازگشت‌پذیری، کدگشایی دوطرفه و مقاومت در برابر خطا می‌باشند.

همچنین چون:

(۱) عملیات کدگذاری پیشنهادی، نوعی نگاشت از کدکلمه‌های شانونی به کدکلمه‌های متقارن بوده و بین آن‌ها یک تناظر یک‌به‌یک و پوشا وجود دارد.

(۲) به ازای هر کدکلمه‌ی شانونی، کدکلمه‌ای هم‌طول با آن در این نگاشت تعریف شده است.

بنابراین می‌توان از (۱) و (۲) نتیجه گرفت که در صورت متقارن‌سازی حجم داده افزایش نخواهد یافت.

در مجموع می‌توان گفت الگوریتم پیشنهادی، یک روش پایه جهت کدگذاری است که روش متفاوتی در رمز کردن داده ها نیز محسوب می‌گردد. در مثال‌هایی که مورد بررسی قرار گرفت حدود ۷ الی ۲۵ درصد فشرده‌سازی در تعداد بیت‌ها بدست آمد. روشن است که این الگوریتم کاربرد فراوانی در تلفیق با سایر روش‌ها خواهد داشت، همانطور که اشاره شد از جمله کاربردهای دیگر آن، متقارن‌سازی بیت‌هاست که به دلیل اهمیت زیاد آن، در مقاله‌ای دیگر به طور مفصل به آن پرداخته خواهد شد. شاید نتیجه‌ی اصلی این‌مقاله توجه به مدیریت داده در سطح بیت‌ها باشد که توجه به نوع چینش و ارتباط بیت‌ها-همانطور که در این‌مقاله نشان داده شد-می‌تواند در جهات مختلف دارای کاربرد باشد. همچنین برای تحقیقات بیشتر توصیه می‌شود به تاثیر جایگشت بیت‌ها بر افزایش بازدهی الگوریتم پیشنهادی پرداخته شود.

## ۷- مراجع

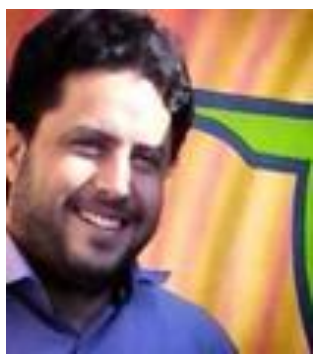

**پرویز قره‌باقری** دانشجوی دکتری ریاضی دانشگاه شاهد است که تحصیلات خود را در رشته‌ی آموزش ریاضی دانشگاه فرهنگیان آغاز کرد و به دنبال آن مدرک کارشناسی ارشد خود را در رشته ریاضی گرایش رمز و کد از دانشگاه شاهد دریافت کرد. همچنین ایشان در مقاطع کارشناسی و کارشناسی ارشد رشته مهندسی نرم افزار نیز به تحصیل و تحقیق پرداخت. وی به عنوان مدرس دانشگاه در رشته‌های کامپیوتر و ریاضی مشغول به تدریس است. زمینه تحقیقاتی وی فشرده‌سازی داده، فشرده‌سازی تصویر، کدگذاری، رمزنگاری و امنیت می‌باشد. آدرس پست الکترونیکی ایشان عبارت است از:

p.gharehbagheri@yahoo.com

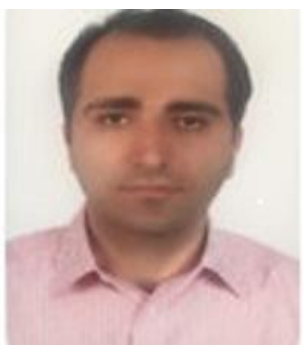

سید حمید حاج سید جوادی مدرک تحصیلی کارشناسی، کارشناسی ارشد و دکتری خود را در دانشگاه صنعتی امیرکبیر، تهران، ایران دریافت نمود. وی به عنوان عضو هیئت علمی تمام وقت و استاد تمام در گروه ریاضیات و علوم کامپیوتر دانشگاه شاهد، تهران، مشغول به کار است. زمینه‌های تحقیقاتی وی جبر کامپیوتر ، شبکه‌های حسگر بیسیم، اینترنت اشیا، رمزنگاری و امنیت است. آدرس پست الکترونیکی ایشان عبارت است از:

h.s.javadi@shahed.ac.ir

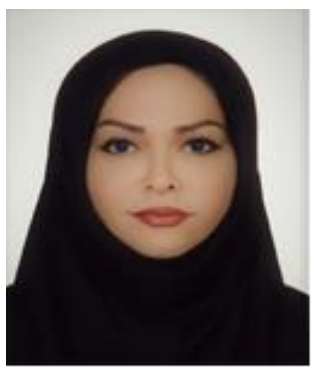

پروانه اصغری عضو هیئت علمی تمام وقت و استادیار گروه مهندسی کامپیوتر دانشگاه آزاد اسلامی واحد تهران مرکزی است. او دوره کارشناسی خود را در رشته مهندسی کامپیوتر نرم افزار از دانشگاه صنعتی شریف، تهران، ایران، کارشناسی ارشد خود در رشته مهندسی کامپیوتر نرم افزار، از دانشگاه علم و صنعت، تهران، ایران و دکترای خود را در رشته مهندسی کامپیوتر از دانشگاه آزاد اسلامی واحد علوم و تحقیقات تهران، ایران به اتمام رساند. زمینه تحقیقاتی وی در حوزه سیستم‌های توزیع شده، اینترنت اشیا، رایانش ابری و محاسبات سرویس گرا است. آدرس پست الکترونیکی ایشان عبارت است از:

p_asghari@iauctb.ac.ir

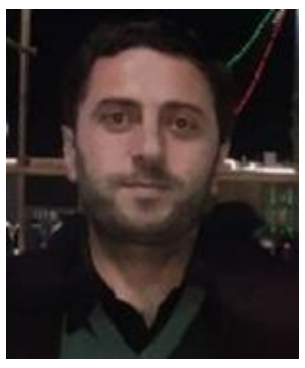

ناصر قره باقری مقطع کارشناسی خود را در رشته ریاضی محض از دانشگاه بیرجند و مقطع کارشناسی ارشد خود را در رشته ریاضی محض گرایش جبرجابجایی از دانشگاه مراغه به پایان رساند. ایشان سال‌هاست که به عنوان مدرس درس ریاضی مشغول به فعالیت می‌باشد. زمینه تحقیقاتی وی جبرجابجایی و گروه‌های خود متشابه می باشد. آدرس پست الکترونیکی ایشان عبارت است از:

ngharehbagheri@yahoo.com

[1] Codeword
[2] Golomb
[3] LZ77